\documentclass[prl,twocolumn,aps,superscriptaddress,showpacs]{revtex4-1}

\usepackage{amsmath,amssymb}
\usepackage{graphicx}
\usepackage{xcolor}
\usepackage{graphicx}
\DeclareGraphicsExtensions{.pdf,.eps,.png,.jpg}
\graphicspath{{./figs/}}
\usepackage{dcolumn}
\usepackage{bm}
\usepackage{tabularx}

\begin{document}

\title{Pressure-induced dimerization and valence bond crystal formation in the Kitaev-Heisenberg magnet $\alpha-$RuCl$_3$}

\author{G. Bastien}
\email{g.bastien@ifw-dresden.de} 
\affiliation{Leibniz-Institut f\"ur Festk\"orper- und Werkstoffforschung (IFW) Dresden, 01171 Dresden, Germany}
\author{G. Garbarino}
\affiliation{European Synchrotron Radiation Facility, 38043 Grenoble, France}
\author{R. Yadav}
\affiliation{Leibniz-Institut f\"ur Festk\"orper- und Werkstoffforschung (IFW) Dresden, 01171 Dresden, Germany}
\author{F. J. Martinez-Casado}
\affiliation{Institut Laue-Langevin, 38042 Grenoble, France}
\affiliation{Instituto de Ciencia de Materiales de Arag\'on (ICMA), CSIC and Universidad de Zaragoza, 50009 Zaragoza, Spain}
\author{R. Beltr\'an Rodr\'iguez}
\affiliation{Leibniz-Institut f\"ur Festk\"orper- und Werkstoffforschung (IFW) Dresden, 01171 Dresden, Germany}
\author{Q. Stahl}
\affiliation{Institut f\"ur Festk\"orper- und Materialphysik, Technische Universit\"at Dresden, 01062 Dresden, Germany}
\author{M. Kusch}
\affiliation{Institut f\"ur Festk\"orper- und Materialphysik, Technische Universit\"at Dresden, 01062 Dresden, Germany}
\author{S. P. Limandri}
\affiliation{IFEG, CONICET, Medina Allende s/n, Ciudad Universitaria, 5000 Cordoba, Argentina}
\author{R. Ray}
\affiliation{Leibniz-Institut f\"ur Festk\"orper- und Werkstoffforschung (IFW) Dresden, 01171 Dresden, Germany}
\affiliation{Dresden Center for Computational Materials Science (DCMS), Technische Universit\"at Dresden, 01062 Dresden, Germany}
\author{P. Lampen-Kelley}
\affiliation{Materials Science and Technology Division, Oak Ridge
National Laboratory, Oak Ridge, Tennessee 37831, USA}
\affiliation{Department of Materials Science and Engineering, University of Tennessee, Knoxville, Tennessee 37996, USA}
\author{D. G. Mandrus}
\affiliation {Materials Science and Technology Division, Oak Ridge
National Laboratory, Oak Ridge, Tennessee 37831, USA}
\affiliation{Department of Materials Science and Engineering,
University of Tennessee, Knoxville, Tennessee 37996, USA}
\author{S. E. Nagler}
\affiliation{Neutron Scattering Division, Oak Ridge National
Laboratory, Oak Ridge, Tennessee 37831, USA}
\author{M. Roslova}
\affiliation{Fakult\"at f\"ur Chemie und Lebensmittelchemie,
Technische Universit\"at Dresden, 01062 Dresden, Germany}
\author{A. Isaeva}
\affiliation{Fakult\"at f\"ur Chemie und Lebensmittelchemie,
Technische Universit\"at Dresden, 01062 Dresden, Germany}
\author{T. Doert}
\affiliation{Fakult\"at f\"ur Chemie und Lebensmittelchemie,
Technische Universit\"at Dresden, 01062 Dresden, Germany}
\author{L. Hozoi}
\affiliation{Leibniz-Institut f\"ur Festk\"orper- und Werkstoffforschung (IFW) Dresden, 01171 Dresden, Germany}
\author{A. U. B. Wolter}
\affiliation{Leibniz-Institut f\"ur Festk\"orper- und Werkstoffforschung (IFW) Dresden, 01171 Dresden, Germany}
\author{B. B\"uchner}
\affiliation{Leibniz-Institut f\"ur Festk\"orper- und Werkstoffforschung (IFW) Dresden, 01171 Dresden, Germany}
\affiliation{Institut f\"ur Festk\"orper- und Materialphysik, Technische Universit\"at Dresden, 01062 Dresden, Germany}
\author{J. Geck}
\affiliation{Institut f\"ur Festk\"orper- und Materialphysik, Technische Universit\"at Dresden, 01062 Dresden, Germany}
\author{J. van den Brink}
\affiliation{Leibniz-Institut f\"ur Festk\"orper- und Werkstoffforschung (IFW) Dresden, 01171 Dresden, Germany}
\affiliation{Institut f\"ur Theoretische Physik, Technische Universit\"at Dresden, 01062 Dresden, Germany}
\affiliation{Department of Physics, Harvard University, Cambridge,
Massachusetts 02138, USA}
\date{\today }

\begin{abstract}
Magnetization and high-resolution x-ray diffraction measurements of the
Kitaev-Heisenberg material $\alpha$-RuCl$_3$ reveal a pressure-induced
crystallographic and magnetic phase transition at a
hydrostatic pressure of $p \sim 0.2$~GPa.
This structural transition into a triclinic phase is characterized by a very strong dimerization of the Ru-Ru bonds, accompanied by a collapse of the magnetic susceptibility. {\it Ab initio} quantum-chemistry calculations disclose a pressure-induced enhancement of the direct
4$d$-4$d$ bonding on particular Ru-Ru links, causing a sharp increase
of the antiferromagnetic exchange interactions. These combined experimental and computational data show that the Kitaev spin liquid phase in $\alpha-$RuCl$_3$ strongly competes with  the crystallization of spin singlets into a valence bond solid.
\end{abstract}


\maketitle

The Kitaev model on a honeycomb lattice has grown into a hot topic
in the last decade due to its exact solubility and its quantum spin-liquid
ground state, which would be relevant for, e.g., quantum
computing~\cite{Knolle2015, Do2017}.  It implies a bond-dependent
 compass-type coupling $K$
 and strong intrinsic spin frustration
\cite{Kitaev2006}. A crucial ingredient for realizing the Kitaev model in real materials is a strong spin-orbit coupling together with a honeycomb structure.
Recently, Kitaev interactions were identified in $\alpha$-RuCl$_3$, from
its unusual magnetic excitation spectrum~\cite{Banerjee2016, Banerjee2017a},
its strong magnetic anisotropy~\cite{Majumder2015},
 and electronic-structure calculations~\cite{Winter16,Yadav2016},
which renders this material an ideal platform for exploring Kitaev magnetism experimentally.

$\alpha$-RuCl$_3$ is a $j_{\mathrm{eff}}$ = 1/2 Mott insulator with a two-dimensional (2D)
layered structure of edge-sharing RuCl$_6$ octahedra forming
a honeycomb lattice. At ambient pressure, the honeycomb layers are
arranged in a monoclinic ($C2/m$) structure at room temperature with one of the three nearest-neighbor (NN) Ru-Ru bonds
slightly shorter than the other two
\cite{cao2016}.
A structural phase transition was reported at $T_S \backsimeq 60$~K under cooling and $T_S \backsimeq 166$~K upon warming, but the low temperature crystal structure is still under debate and could be either rhombohedral ($R\bar{3}$) ~\cite{Park2016, Reschke2018} or monoclinic ($C2/m$) \cite{Little2017, Lampen-Kelley2018}.
The onset of long-range magnetic order at $T_N \backsimeq 7$~K~\cite{cao2016} in $\alpha$-RuCl$_3$ implies that
other magnetic interactions have to be considered in addition
to the Kitaev interaction $K$: a NN Heisenberg $J$, an off-diagonal coupling $\Gamma$, as well as
next-NN interactions $J_2$ and $J_3$~\cite{Winter16,Yadav2016,Janssen2017,Winter2018}. 
While electronic-structure calculations indicate that $K$ is ferromagnetic in $\alpha$-RuCl$_3$
and indeed defines the largest exchange energy scale~\cite{Winter16,Yadav2016,Janssen2017,Winter2018}, the debate on the minimal effective spin model
and precise magnitude of the different couplings is not fully settled yet.
 By applying a
magnetic field in the basal plane, the magnetic zigzag ground state can be suppressed~\cite{Majumder2015, Sears2017,Wolter2017} and the phase
above this transition was identified as a quantum spin liquid,
by NMR~\cite{Baek2017}, thermal conductivity~\cite{Leahy2017,Yu2018, Hentrich2018}, Terahertz spectroscopy~\cite{Wang2017a} and neutron scattering experiments~\cite{Banerjee2018}.

Further, it was very recently
shown by specific heat, magnetization and NMR measurements~\cite{Wang2017, Cui2017} that the
N\'eel temperature of $\alpha$-RuCl$_3$ increases slightly with pressure and vanishes through a phase separation regime around 0.5 GPa at finite temperature. Thermal expansion measurements at ambient pressure predicted also the suppression of the magnetic order under pressure~\cite{He2017}. However, the initial slope value ${dT_N}/{dp}_{p=0} \backsimeq -23$~K/GPa from thermal expansion is in contradiction with the phase diagram drawn by the other techniques under the application of hydrostatic pressure~\cite{Wang2017, Cui2017}. Magnetization measurements indicate a reduction of the in-plane magnetization and a high-temperature transition of unknown origin, while NMR indicates no long-range magnetic order and gapped magnetic fluctuations in the high-pressure state~\cite{Cui2017}. Furthermore, electrical resistivity studies under hydrostatic pressure exclude the possibility of a pressure-induced insulator-to-metal transition~\cite{Wang2017}. 

To clarify the nature of this pressure-induced phase we bring together three essential pieces of information:  detailed magnetization and x-ray diffraction measurements on $\alpha$-RuCl$_3$ under
hydrostatic pressure which are combined with a set of quantum chemistry electronic-structure
calculations.
Together they unequivocally evidence that pressure induces a first-order structural transition
from the rather regular Kitaev-Heisenberg honeycomb system towards
a pronounced nonmagnetic dimer state with a large difference between the long and the short Ru-Ru distance of about 0.7~\AA.
 {\it Ab initio} computations for the high-pressure  crystal structures reveal
remarkably large isotropic antiferromagnetic couplings on the short Ru-Ru bonds, in the range
of hundreds of meV, which explain the experimentally observed nonmagnetic state of $\alpha$-RuCl$_3$.
We show that the  $j_{\mathrm{eff}}$ = 1/2 picture is significantly modified under
hydrostatic pressure as a result of a reduction of spin-orbit-coupling effects due to
increased crystal-field splittings in the high-pressure phase.

$\alpha$-RuCl$_3$ single crystals were grown from phase-pure
commercial RuCl$_3$ powder via a high-temperature vapor transport
technique~\cite{Banerjee2016,Hentrich2018,note1}. Both magnetization and x-ray diffraction
(XRD) show the homogeneous high-quality nature of our single crystals. 

Magnetization under hydrostatic pressure was measured in a home-built pressure cell
for a commercial superconducting quantum interference device (SQUID) magnetometer from Quantum Design. 
Two opposing, conical ceramic anvils compress a CuBe gasket with
a small hole that serves as a sample chamber~\cite{Tateiwa2011}.
Daphne oil 7373 is used as a pressure transmitting medium ensuring
good hydrostatic conditions up to about 2~GPa. The pressure was applied at room temperature and
determined at $T \backsimeq$ 7~K from the superconducting transition of a lead sample. 
The magnetic response for the empty cell was measured separately and subtracted from the data, in order to achieve an accuracy on the absolute value of the magnetization of about $2.10^{-3}$~emu/mol/Oe at $\mu_0H = 1~$T.

X-ray diffraction (XRD) experiments as a function of temperature down to 30~K and pressure up to 11~GPa were performed at the beamline ID27 of the European Synchrotron Radiation Facility, using a monochromatic
beam with a photon energy of 33~keV focused down to a
spot size of 3x3 $\mu m^{2}$. High-quality single crystals
were loaded into a membrane-driven diamond anvil cell (DAC) filled with helium as the
pressure-transmitting medium. The DAC assembly was then mounted in
a continuous He-flow cryostat, allowing one to cool the sample while
continuously monitoring the pressure in the sample space via ruby
fluorescence. Additional experiments at ID27 without a DAC, i.e.
at ambient pressure, were performed as well. The collected three-dimensional data were integrated and corrected for Lorentz-,
polarization and background effects using the CrysAlis$^{Pro}$
software~\cite{Crys2016}. The subsequent weighted full-matrix least-squares refinements on $F^2$ were done with SHELX-2012~\cite{Sheldrick2007} as implemented in the WinGx 2014.1 program suite~\cite{Farrugia1999}.

\begin{figure}[!h]
\begin{center}
\includegraphics[width=0.8\linewidth]{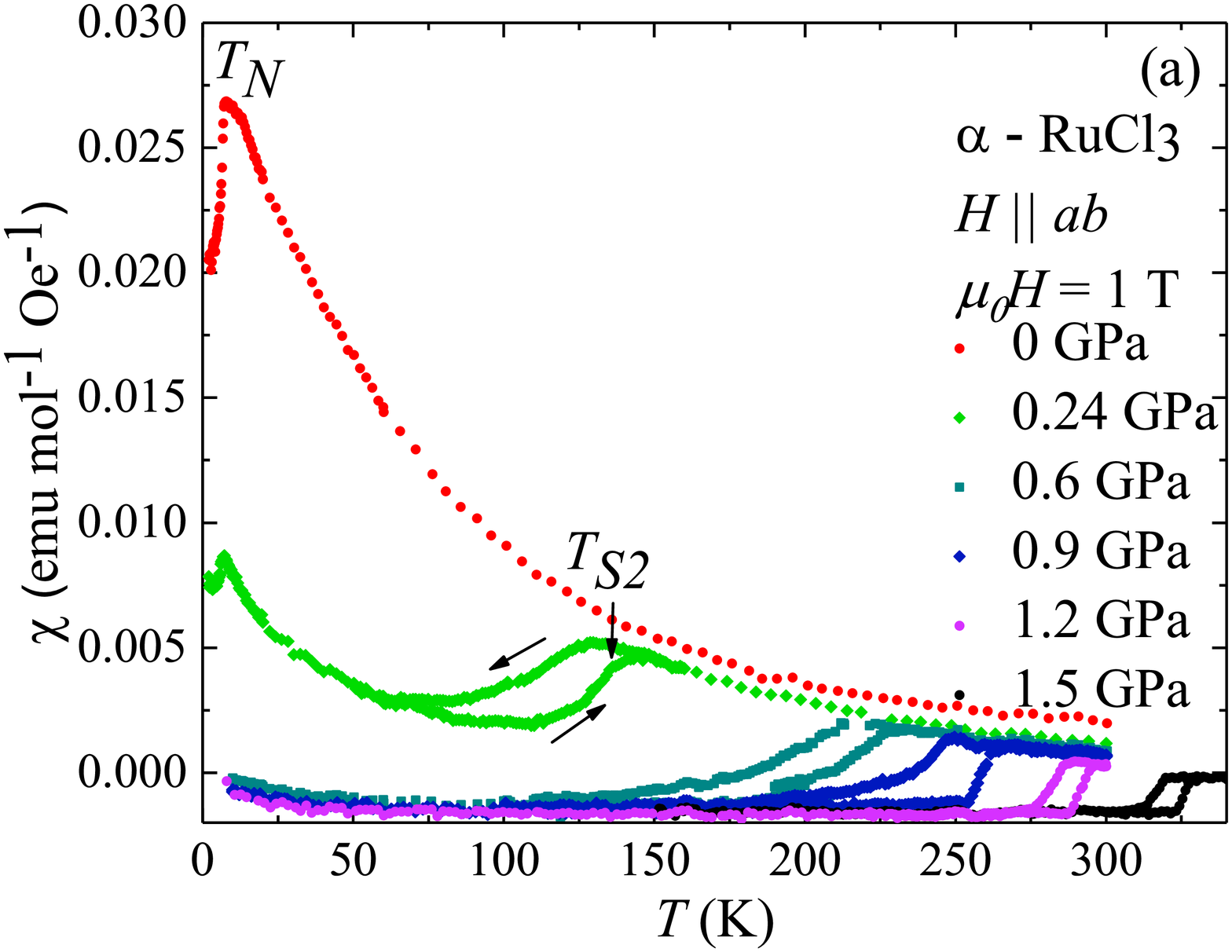}
\includegraphics[width=0.77\linewidth]{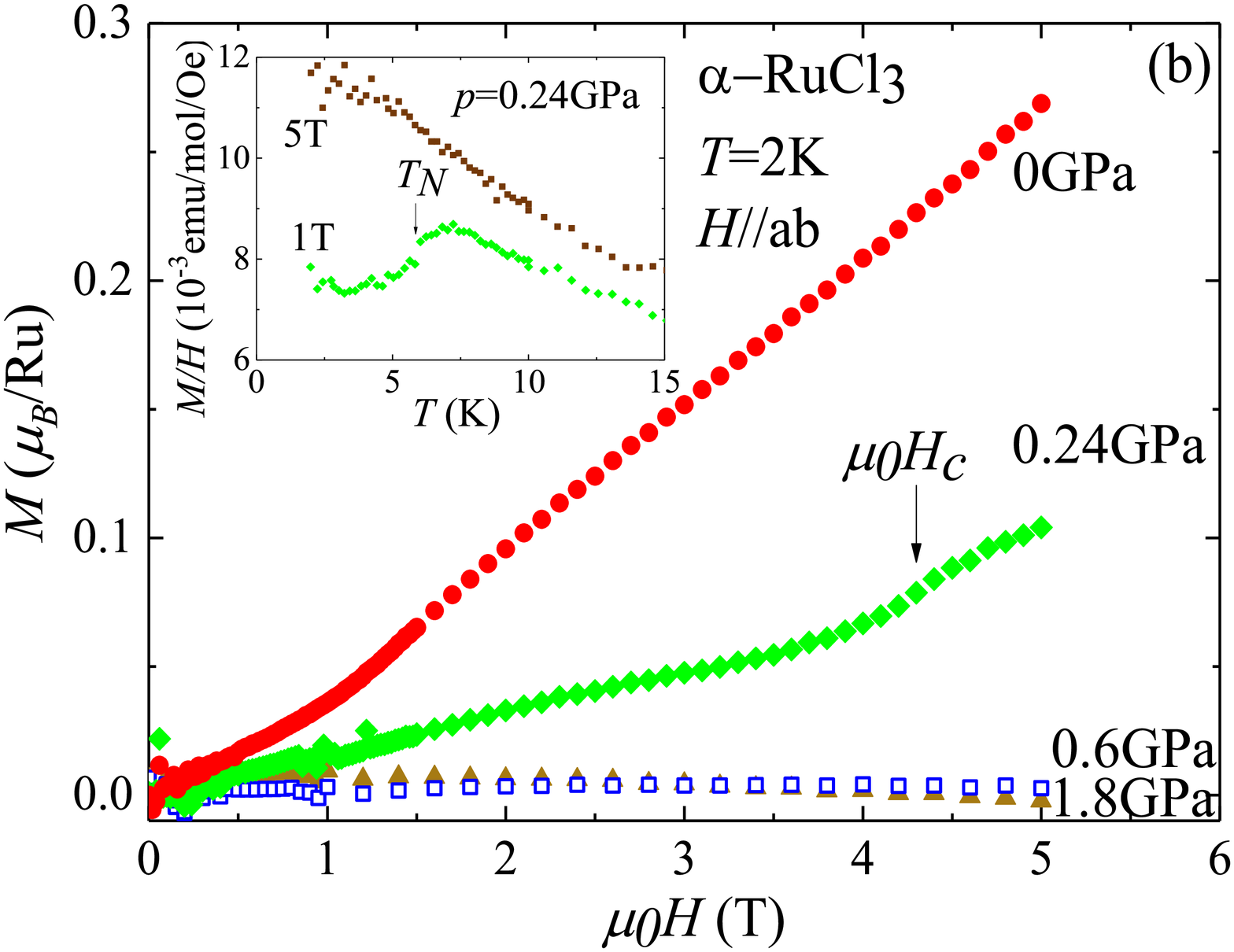}
\caption{(a) Field-cooled magnetic susceptibility of
$\alpha$-RuCl$_3$ as a function of temperature for different
pressures. A magnetic field of 1~T was applied in the $ab$ plane.
The structural transition $T_{S2}$ shows thermal hysteresis, and the
cooling and warming curves are indicated by black arrows around
the 0.24~GPa curve. (b) Magnetization of $\alpha$-RuCl$_3$ at 2~K
as a function of magnetic field applied in the $ab$ plane for
different pressures. $H_c$ indicates the phase transition from the
zigzag order towards the field-induced quantum
spin liquid.
The inset shows the renormalized magnetization $M/H$ at $p = 0.24$~GPa
as a function of temperature for magnetic fields of $\mu_0H$ = 1~T
and $\mu_0H$ = 5~T. $T_N$ indicates the magnetic phase transition
from the zigzag order to the paramagnetic state.} 
\label{MvsTp}
\end{center}
\end{figure}

\begin{figure}[t]
\begin{center}
\includegraphics[width=0.8\linewidth]{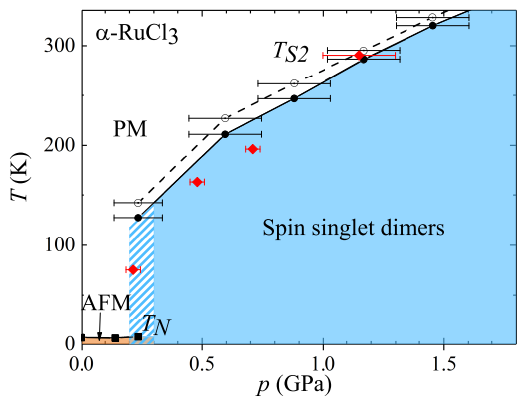}
\caption{Temperature-pressure phase diagram of $\alpha$-RuCl$_3$. The solid and open black circles represent the transition $T_{S2}$ in magnetization by cooling and by warming, respectively. The red squares represent the transition $T_{S2}$ from x-ray diffraction. The striped area represents the region where phase separation
occurs. The error bars on pressure for the magnetization measurements come from the thermal expansion of the pressure cell. The lines are guides to the eye.}
\label{pT}
\end{center}
\end{figure}

The magnetic susceptibility $\chi$ of $\alpha$-RuCl$_3$ in the
$ab$ plane is presented in Fig.~\ref{MvsTp} as a function of $T$
and for different pressures up to 1.5~GPa. At ambient pressure the
antiferromagnetic transition into the zigzag ordered ground state
is clearly observable at $T_N\backsimeq 7$~K. Under a small
hydrostatic pressure of 0.24~GPa, a second phase transition occurs
at $T_{S2} = 140$~K with a reduction of the magnetic susceptibility
by about a factor of two. This transition shows thermal hysteresis,
indicating its first-order structural nature. At 0.6~GPa and
higher pressures the transition is shifted to higher temperatures,
followed by a strong suppression of the magnetic susceptibility
below $T_{S2}$. Note that the measurement at 0.24~GPa can be interpreted as a phase separation
in the sample: While a fraction of the sample is in the
high-pressure state and yields a magnetization close to zero below
$T_{S2}$, the second fraction still shows a paramagnetic behavior
down to $T_N \backsimeq 7$~K, followed by an antiferromagnetic
zigzag state below $T_N$. These results are qualitatively in good agreement with independent magnetization measurements~\cite{Cui2017}.
While the authors in Ref.~\cite{Cui2017} measured the magnetic susceptibility on a single crystal with a single N\'eel temperature $T_N \backsimeq 13.6$~K, indicating an AB stacking of the honeycomb layers~\cite{cao2016}, the measurements reported here were performed on a single crystal with an ABC stacking as indicated by $T_N \backsimeq 7$~K~\cite{cao2016}. The pressure induced collapse of the in plane magnetization in Ref.~\cite{Cui2017} with an AB stacking seems shifted to higher pressure compared to the one reported here and shows a phase separation regime on a broader pressure range up to at least $p \backsimeq 1~$GPa. This difference suggests that the stacking sequence would have a small influence on the structural transition $T_{S2}$.

In order to obtain a deeper insight into the pressure-induced
magnetic ground state of $\alpha$-RuCl$_3$, we performed additional
measurements of $\chi$ along the
transverse axis $c^\ast$ under hydrostatic pressure, which confirm the collapse of the
magnetic susceptibility below $T_{S2}$ and thus the nonmagnetic
nature of the high-pressure state of $\alpha$-RuCl$_3$ (cf.
Supplemental Material~\cite{supplement}).

The magnetization at 2~K as a function of the magnetic field
applied in the basal plane is represented in Fig.~\ref{MvsTp}(b). The magnetization at $p= $0.24~GPa shows an upward step at $\mu_0H_c = 4.3$~T. Since the temperature scan at 5~T represented in the inset of Fig.~\ref{MvsTp}(b) confirms the absence of the
antiferromagnetic transition above $H_c$, the critical field $H_c$ corresponds to the suppression of the zigzag order by an
external magnetic field similar to $\mu_0H_c \backsimeq $7-8~T at
ambient pressure~\cite{Majumder2015, Wolter2017,Baek2017}. Thus,
the quantum critical point toward the field-induced quantum spin-liquid
state seems to be strongly reduced from its ambient
pressure value in this regime. At even higher pressures of 0.6
and 1.8~GPa a collapse of the magnetic signal is observed in the
(pure) high-pressure state up to 5~T, preventing any
extraction of the magnetic susceptibility on an absolute scale
within the accuracy of our experimental setup.

The resulting temperature-pressure ($T-p$) phase diagram of
$\alpha$-RuCl$_3$ is given in Fig.~\ref{pT}. The N\'eel temperature stays rather constant up to about 0.2~GPa. Then, $\alpha$-RuCl$_3$ undergoes a pressure-induced phase transition
around 0.2~GPa into a nonmagnetic state with phase separation occurring over
a finite pressure range. The transition temperature $T_{S2}$
increases rapidly with pressure and reaches room temperature
around $p$ = 1.3~GPa. This phase diagram is in good agreement with previous studies under hydrostatic pressure~\cite{Wang2017, Cui2017} and further shows that the pressure-induced transition is of a first-order nature.

\begin{figure*}[t!]
    \includegraphics[width=0.6\textheight]{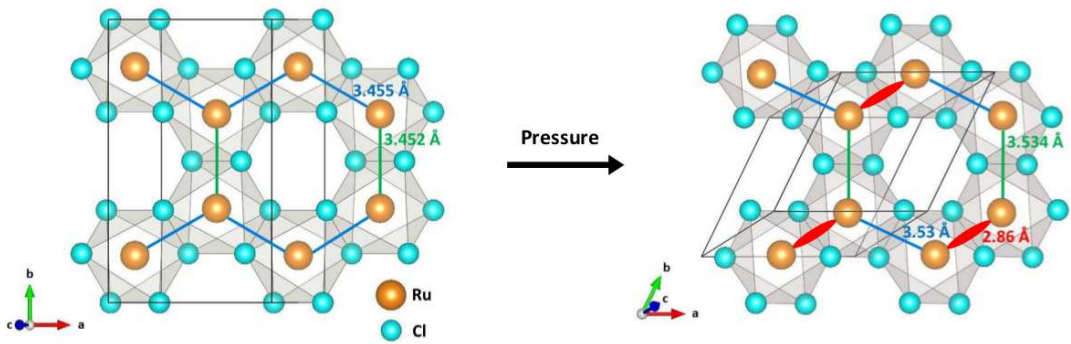}
        \caption{Honeycomb layer of the $\alpha$-RuCl$_3$ structure at 300~K in the monoclinic phase at ambient pressure (left) and in the triclinic phase at 2.08~GPa (right). The ellipses represent the pressure-induced Ru-Ru dimers.}
      \label{fig:Structure}
\end{figure*}

In order to elucidate the microscopic mechanisms underlying the pressure-induced
transition at $T_{S2}$, we performed high-resolution XRD under
hydrostatic pressure. At ambient pressure and ambient temperature
our XRD measurements confirm the monoclinic $C2/m$ structure reported
earlier~\cite{cao2016, supplement}. Upon
increasing pressure, however, a transition $T_{S2}$ into a triclinic
$P\bar{1}$ phase with Ru-Ru dimers was observed together with the
changes observed in the magnetic susceptibility as shown in the phase diagram given in Fig.~\ref{pT}. The slight difference between the points from magnetization and from x-ray diffraction in this phase diagram can be explained by uncertainties regarding the pressure of the magnetization measurements, by finite transition widths, and by small sample dependencies~\cite{note1}. The triclinic phase with Ru-Ru dimers is stable up to the highest applied pressure of 11~GPa.

In order to determine the structural changes in more detail, we performed refinements of the measured intensities at various pressures. The obtained structural changes upon entering the triclinic high-pressure phase are illustrated in
Fig.~\ref{fig:Structure}.
Besides changes in the relative positions of neighboring RuCl$_3$-layers, there are dramatic changes within the layers themselves.
At ambient
pressure (left panel of  Fig.~\ref{fig:Structure}), 
the differences in the Ru-Ru distances are only about 0.003~\AA (Table 1 of Supplemental Material~\cite{supplement}), resulting in a
nearly hexagonal honeycomb lattice. The transformation into the
triclinic phase with increasing pressure involves the formation of
Ru-Ru dimers with a large difference between the short and the long Ru-Ru
distances of about 0.7~\AA. This extremely strong dimerization involves all Ru atoms,
i.e., every Ru atom is part of a dimer.


\begin{figure}[t]
\centering
\includegraphics[width=6.1cm]{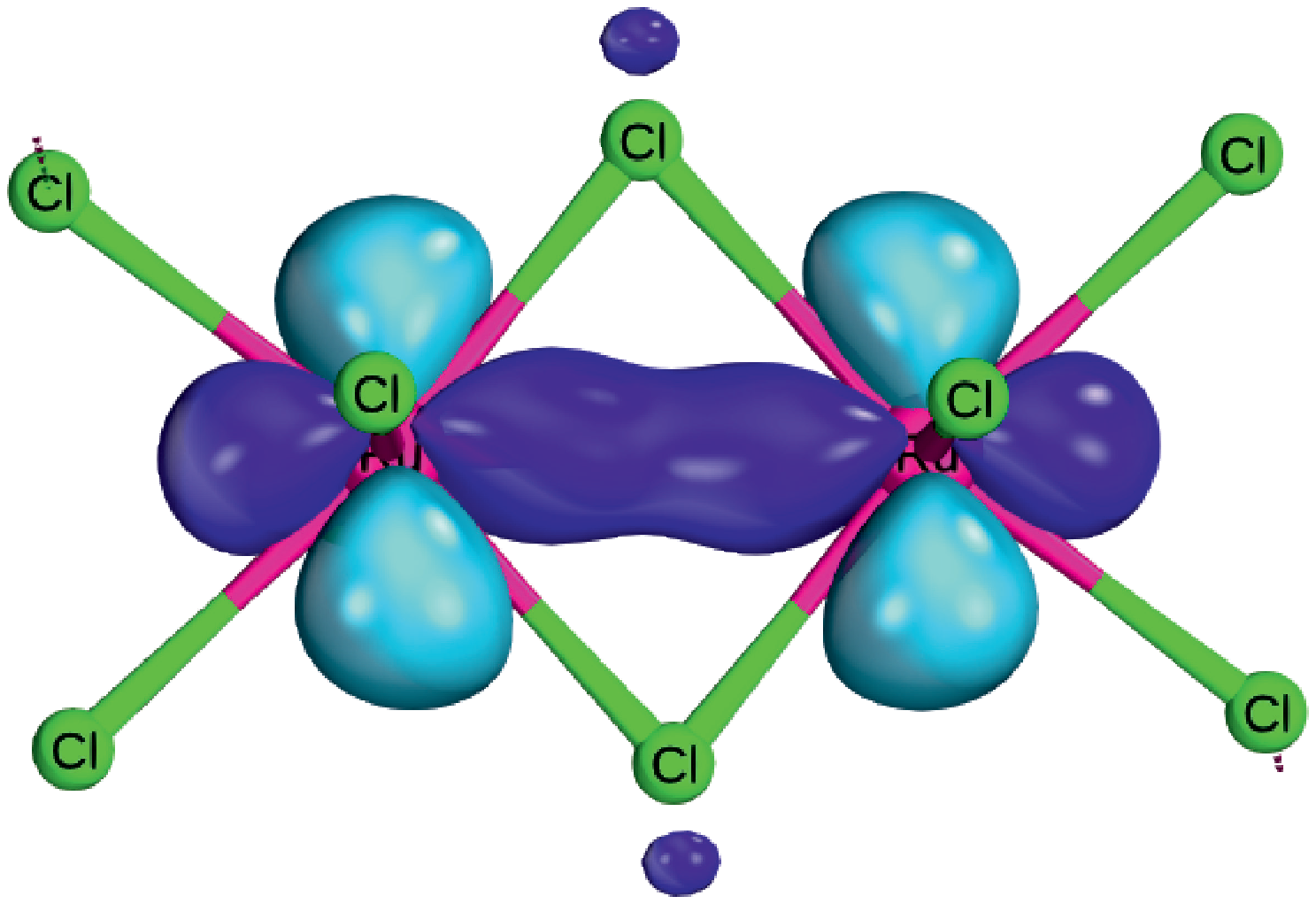}
\includegraphics[width=6.1cm]{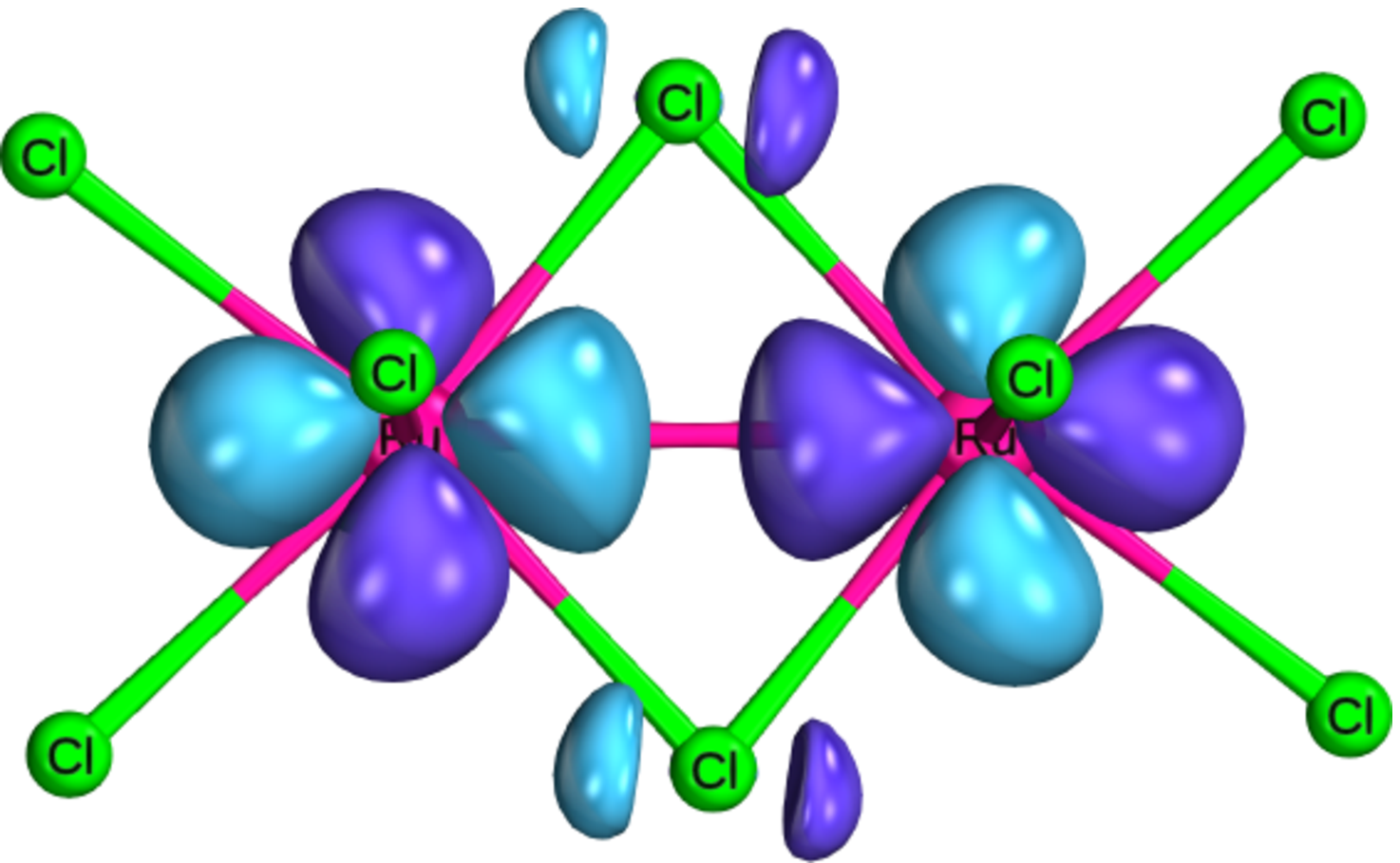}
\caption{
Bonding (top) and antibonding (bottom) combinations of the Ru $t_{2g}$ hole orbitals on the shorter
Ru-Ru bonds of the crystal structure in the dimer state, as obtained by embedded-cluster quantum chemistry
calculations.
}
\label{fig:overlap}
\end{figure}

\begin{table}[b]
\caption{
NN magnetic couplings (meV) for high-pressure crystal structures as determined at room temperature;
results of spin-orbit MRCI calculations for the longer Ru-Ru links, where the isotropic and
anisotropic components still have comparable strength.
}
\begin{center}
\label{RuCl3:tab}
\begin{tabular}{cccccc}
 \hline \hline
     Pressure (GPa)  &  \hspace{0.2cm}  $K$  &\hspace{0.2cm} $J$ &\hspace{0.2cm}  $\Gamma_{xy}$   &\hspace{0.2cm}  $\Gamma_{zx} = -\Gamma_{yz}$  \\
      \hline
      \vspace{-0.2cm}\\
     $ 4.60$    & \hspace{0.2cm} $-3.15$   &\hspace{0.2cm} $3.32$  & \hspace{0.2cm} $-0.22$  &  $-0.95$ \\
         \vspace{-0.5cm}\\
               \vspace{-0.1cm}\\
     $10.60$  & \hspace{0.2cm} $-1.75$   &\hspace{0.2cm} $0.81$  &\hspace{0.2cm}  $0.80$  &  $-0.49$ \\
         \vspace{-0.5cm}\\
           \vspace{-0.1cm}\\
         \hline
         \hline
 \end{tabular}
\end{center}
\end{table}

To clarify the effect of this dimerization on magnetism, we carried out embedded-cluster quantum-chemistry calculations using the experimental crystal structures.
Our {\it ab initio} results show that in the triclinic phase 
the $4d$-shell $t_{2g}$
crystal-field
splittings are very large,
up to 0.35 eV,
and counteract the effect of spin-orbit coupling.
The $j_{\rm eff}$=1/2 picture is
therefore significantly modified and
given the peculiar character of the Ru $t_{2g}$ hole, an antiferromagnetic isotropic spin model turns out to be
a rather good approximation on the shortest
Ru-Ru links, with an impressively strong antiferromagnetic exchange.
Specifically, we find that two of the $4d$ $t_{2g}$ levels are nearly degenerate, lie at lower
energy (electron picture), and that the $t_{2g}$ hole is mainly associated with the high-energy orbital that provides
a large direct $d$-$d$ overlap on the shortest Ru-Ru bonds, as sketched in Fig.~\ref{fig:overlap}.
By multireference configuration-interaction (MRCI) calculations~\cite{Helgaker2000} we derive 
singlet-triplet separation energies as high as 440 and 550 meV for the shortest Ru-Ru links found experimentally at 300~K for 4.6 and 10.6 GPa,
respectively, with vanishing splittings among the triplet components.
So large energy differences between the singlet and triplet states associated with two NN $t_{2g}^5$ ions imply that a finite magnetization can only be achieved by
very large magnetic fields, which is indeed observed in
Fig.~\ref{MvsTp}(b), and also explain the large spin-excitation gap observed in a recent NMR study of $\alpha$-RuCl$_3$ under  pressure~\cite{Cui2017}.

For the longer Ru-Ru links, the relevant effective model is an extended
pseudospin-$1/2$
Hamiltonian with both isotropic and anisotropic components~\cite{Yadav2016}\,,
\begin{equation}
 {\cal H}_{i,j} =J\, \tilde{\bf{S}}_i \cdot \tilde{\bf{S}}_j
           +K \tilde{S}^z_i \tilde{S}^z_j
           +\sum_{\alpha \neq \beta} \Gamma_{\!\alpha\beta}(\tilde{S}^\alpha_i\tilde{S}^\beta_j +
                                                          \tilde{S}^\beta_i \tilde{S}^\alpha_j), \ \
\label{Eq:ham}
\end{equation}
where $\tilde{\bf{S}}_i$ and $\tilde{\bf{S}}_j$ 
are
NN
pseudospin 1/2 operators
and the $\Gamma_{\alpha\beta}$ coefficients 
stand for
off-diagonal
couplings
of the anisotropic exchange
tensor
with $\alpha,\beta \in \{x,y,z\}$.
Mapping the
spin-orbit MRCI results
onto such an effective model
 \cite{Yadav2016,Bogdanov15},
we arrive at the NN couplings listed in Table~\ref{RuCl3:tab} for the long links of the dimerized structure at 4.6~GPa and 10.6~GPa.

The combined experimental and theoretical results therefore reveal
a competition between spin-orbit coupling and covalency effects.
Interestingly, a second crystal structure, $\beta$-RuCl$_3$, with Ru chains at room temperature and Ru dimers at low temperature was reported~\cite{Hillebrecht2004}.
Previous density-functional calculations predicted that
$\alpha$-RuCl$_3$ would also dimerize in 
the
absence of spin-orbit coupling~\cite{Kim2016}.
While at ambient pressure the spin-orbit coupling is significantly larger than the crystal-field splittings 
to stabilize
a Ru $j_{\mathrm{eff}} \backsimeq 1/2$ state,
with increasing pressure a phase dominated by strong covalency appears~\cite{Jackeli2007, Jackeli2008, Streltsov2016}.
This mechanism may also apply to other $4d$ and $5d$ metal halides and oxides such as $\alpha$-MoCl$_3$ at ambient pressure~\cite{Schaefer1967, McGuire2017} and the Kitaev-Heisenberg iridate $\alpha$-Li$_2$IrO$_3$ at a critical pressure $p_c \backsimeq 3.8$~GPa~\cite{Hermann2018}. Thus dimerization may be a rather general feature of $4d$ and $5d$ honeycomb systems, due to a subtle interplay between spin-orbit coupling, intermetallic bonding, and magnetism.

In summary, our magnetization and x-ray diffraction experiments on $\alpha$-RuCl$_3$ under hydrostatic pressure
show a pressure-induced phase transition
from the monoclinic to a triclinic structure, featuring a very pronounced Ru-Ru dimerization and a valence bond crystal of ordered dimers. The latter are characterized by remarkably strong antiferromagnetic isotropic couplings due to an increased direct overlap of the
 Ru $4d$ $t_{2g}$ orbitals.
This dimerization leads to a complete suppression of the magnetization and thus to a
 pressure-induced nonmagnetic
state of $\alpha$-RuCl$_3$.
 Our
results show that the Kitaev physics 
in this $d$-electron honeycomb system is in competition with the formation of spin singlet valence bonds: 
Indeed, $\alpha$-RuCl$_3$ shows the occurrence of both a quantum spin liquid state under magnetic field, which is relevant for its topological properties, and a spin solid under hydrostatic pressure: the spin singlet valence bond crystal. Thus, this material will provide new insights for the study of concomitant magnetic and lattice instabilities in 4$d$ and 5$d$ metal halides and oxides.

\begin{acknowledgments}
We acknowledge insightful discussions with Manuel Richter, Matthias Vojta, Lukas
Janssen, and Hans-Henning Klauss. This research has been supported
by the DFG via SFB 1143 and WO 1532/3-2 and by the European Union's Horizon 2020 research and innovation programme under the Marie Sklodowska-Curie Grant Agreement No. 796048. M. Mezouar is acknowledged for providing beamtime at ID27 at the ESRF and for fruitful discussions. S.P.L. is grateful to CONICET for financial support during her stay at the ESRF. R.Y. and L.H. acknowledge Ulrike Nitzsche for technical support as concerns the {\it ab initio} calculations. P.L.K and D.G.M. were supported by the Gordon and Betty Moore Foundation EPiQS Initiative Grant No. GBMF4416. S.N. was supported by the US DOE, BES, Scientific User Facilities Division under contract DE-AC0500OR22725
with the Oak Ridge National Laboratory.
\end{acknowledgments}


\begin{thebibliography}{42}
\bibitem [{not()}]{note1}%
  \BibitemOpen
  \href@noop {} {}\bibinfo {note} {The single crystals for the magnetization
  studies were grown at the Oak Ridge National Laboratory, while the single
  crystals for the x-ray diffraction experiments were grown at the TU
  Dresden.}\BibitemShut {Stop}%
\end{thebibliography}

%
\end{document}